\documentclass[superscriptaddress,twocolumn]{revtex4-1}
\usepackage{amssymb}
\usepackage{amsmath}
\usepackage{cases}
\usepackage{CJK}
\usepackage{graphicx}
\usepackage{dcolumn}
\usepackage{bm}

\begin{document}

\title{Self-bound dipolar quantum droplets in an optical lattice}
\author{Yu-Jia Gao}
\affiliation{College of Physics and Electronic Information Engineering, Qinghai Normal
University, Xining, Qinghai 810016, China}
\author{La-Tai You}
\affiliation{College of Physics and Electronic Information Engineering, Qinghai Normal
University, Xining, Qinghai 810016, China}
\author{Gong-Ping Zheng}
\affiliation{College of Physics and Electronic Information Engineering, Qinghai Normal
University, Xining, Qinghai 810016, China}
\affiliation{Academy of Plateau Science and Sustainability, Xining, Qinghai
810016, China}
\date{\today }

\begin{abstract}
A self-bound dipolar quantum droplet modulated by an shallow optical lattice
are studied. It is found that the behaviors of the droplets in an optical
lattice are similar with those without the optical lattice. In the shallow
enough limit the properties of the periodically-modulated dipolar droplets
are exactly the same with those without the optical lattice, which are
self-bound for the absence of any trap in the polarization direction. So we
conclude that the formation of droplets in an shallow optical lattice is the
competition of atomic interactions and the quantum fluctuations. The optical
lattice can only modulate the droplets. Our model can also describe the
finite-size dipolar droplet system.
\end{abstract}

\maketitle

\section{Introduction}

Quantum droplet, which behaves as liquid but with density being orders of
magnitude thinner than air \cite{fer19}, has become an active field in both
experimental and theoretical research, see the review papers \cite%
{pfau16,pfau21}. Quantum droplet was firstly suggested in a two-species
condensate, stabilized by quantum fluctuations \cite{petrov15,petrov17}. A
quantum droplet was first observed in the experiment of dipolar $^{164}Dy$
atoms \cite{pfau16n1,pfau16n2}, which have large magnetic moment. Soon
after, the quantum droplet was also observed in the experiment of
two-species condensate \cite{cab17,sem18}\textit{.}

The dipole-dipole interaction of dipolar atoms is long-ranged and
anisotropic, which is repulsive (attractive) if the dipoles are arranged
side-by-side (head-to-tail), depending on the aspect ratio of the external
harmonic trap. The dipolar condensate may become unstable when the
attractive dipole-dipole interaction overwhelms the repulsive and isotropic
contact-interaction \cite{pfau09}, which results in the growth of atom
density and three-body loss and eventually the collaps of condensate. While
the dense atom-cloud could be stabilized by quantum fluctuations when the
two competing interactions are close to each other, where the quantum
fluctuations dominant. Because the energy of quantum fluctuations grows
faster than others with the increase of atom number \cite{pfau16,pfau21},
thus the droplet state was formed when the energy of quantum fluctuations
increase to a critical value.

The dipolar droplets with or without the harmonic trap have been extensively
investigated both theoretically and experimentally \cite%
{santos161,santos162,santos163,blak161,blak162,blak163}. The dipolar droplet
in an optical lattice have also been studied very recently \cite%
{santos20,santos23,mor23}. But the existing results focus on the deep
optical lattice, where the atoms are localized at the bottom of the lattice.
The system is described by the well-known Bose-Hubbard model with the
wavefunction expanding by the local Wannier functions. In the present paper
we study the properties of a self-bound dipolar droplet modulated by a
shallow optical lattice. A periodically-modulated dipolar quantum droplet is
observed. The quantum droplets of two-species condensates in a deep optical
lattice have been studied recently \cite{mor20,mor21,machid22,va24}. The
quantum droplets on a shallow optical lattice have also been studied very
recently \cite{nie23,kar24}. But they considered only the quantum droplets
of two-species condensates.

The paper is organized as follows: In Sec. II, we introduce the model of a
three dimensional dipolar condensate in a composite trapping potential. In
Sec. III, we consider the properties of self-bound droplets without an
optical lattice and the effects of an optical lattice on the properties of
the self-bound droplets is studied in Sec. IV. Finally, we summarize our
results and give some remarks in Sec. V.

\section{Model}

We consider a trapped dipolar condensate with the quantum fluctuations \cite%
{pfau16,pfau21}, which the energy functional is

\begin{eqnarray}
E\left[ \psi \right] &=&N\int d\mathbf{r}\left[ \frac{\hbar ^{2}}{2M}%
\left\vert \mathbf{\triangledown }\psi \right\vert ^{2}+V\left( \mathbf{r}%
\right) \left\vert \psi \right\vert ^{2}+\frac{g}{2}N\left\vert \psi
\right\vert ^{4}\right.  \label{ef} \\
&&\left. +\frac{1}{2}\left\vert \psi \right\vert ^{2}\Phi _{dd}+\frac{2}{5}%
g_{\text{qf}}\left\vert \psi \right\vert ^{5}\right] ,  \notag
\end{eqnarray}%
where $\psi $ is the order parameter of the condensate, which is normalized
to unit here. The parameter $g=4\pi \hbar ^{2}a_{s}/M$ describes the contact
interaction with the $s$-wave scattering length $a_{s}$ and the mass of
dipolar atoms $M$, which are chosen as the parameters for $^{164}Dy$ atoms
\cite{pfau16,pfau21} in this paper. $N$ is the number of condensed atoms.

The effective dipole-dipole interaction potential is

\begin{equation}
\Phi _{dd}=N\int d\mathbf{r}^{^{\prime }}\left\vert \psi \left( \mathbf{r}%
^{^{\prime }},t\right) \right\vert ^{2}U_{dd}\left( \mathbf{r-r}^{^{\prime
}}\right)
\end{equation}%
with the two-body dipole-dipole interaction

\begin{equation}
U_{dd}\left( \mathbf{r-r}^{^{\prime }}\right) =\frac{\mu _{0}\mu _{m}^{2}}{%
4\pi }\frac{1-3\text{cos}^{2}\theta }{\left\vert \mathbf{r-r}^{^{\prime
}}\right\vert ^{3}},
\end{equation}%
where $\mathbf{r-r}^{^{\prime }}$ is the relative position of the two
interacting dipoles. $\theta $ is the angle between the relative position
and the direction of polarization $z$. $\mu _{0}$ is the vacuum permittivity
and $\mu _{m}=10\mu _{B}$ is the dipole moment of $^{164}Dy$ atom, where $%
\mu _{B}$ is the Bohr magneton.

The last term in equation (\ref{ef}) is the beyond mean-field correction
stemming from quantum fluctuations. For the dipolar condensate, this
correction is given by \cite{lima11,lima12}

\begin{equation}
g_{\text{qf}}=\frac{32}{3\sqrt{\pi }}gN^{3/2}\sqrt{a_{s}^{3}}Q_{5}\left(
\varepsilon _{dd}\right) ,
\end{equation}%
where $Q_{5}=\frac{1}{2}$ $\int_{0}^{\pi }d\theta \sin \theta \left[
1+\varepsilon _{dd}\left( 3\cos ^{2}\theta -1\right) \right] ^{5/2}$
describes the averged angular contribution of the dipolar interaction. The
strength of quantum fluctuations depends on the ratio of the interaction
length scales $\varepsilon _{dd}=a_{dd}/a_{s}$, where $a_{dd}=M\mu _{0}\mu
_{m}^{2}/\left( 12\pi \hbar ^{2}\right) $ is the length scale of the dipolar
interaction and is about $131a_{B}$ for $^{164}Dy$ atoms, with $a_{B}$ the
Bohr radius.

\begin{figure}[tbp]
\centering\includegraphics  [width=8.6cm] {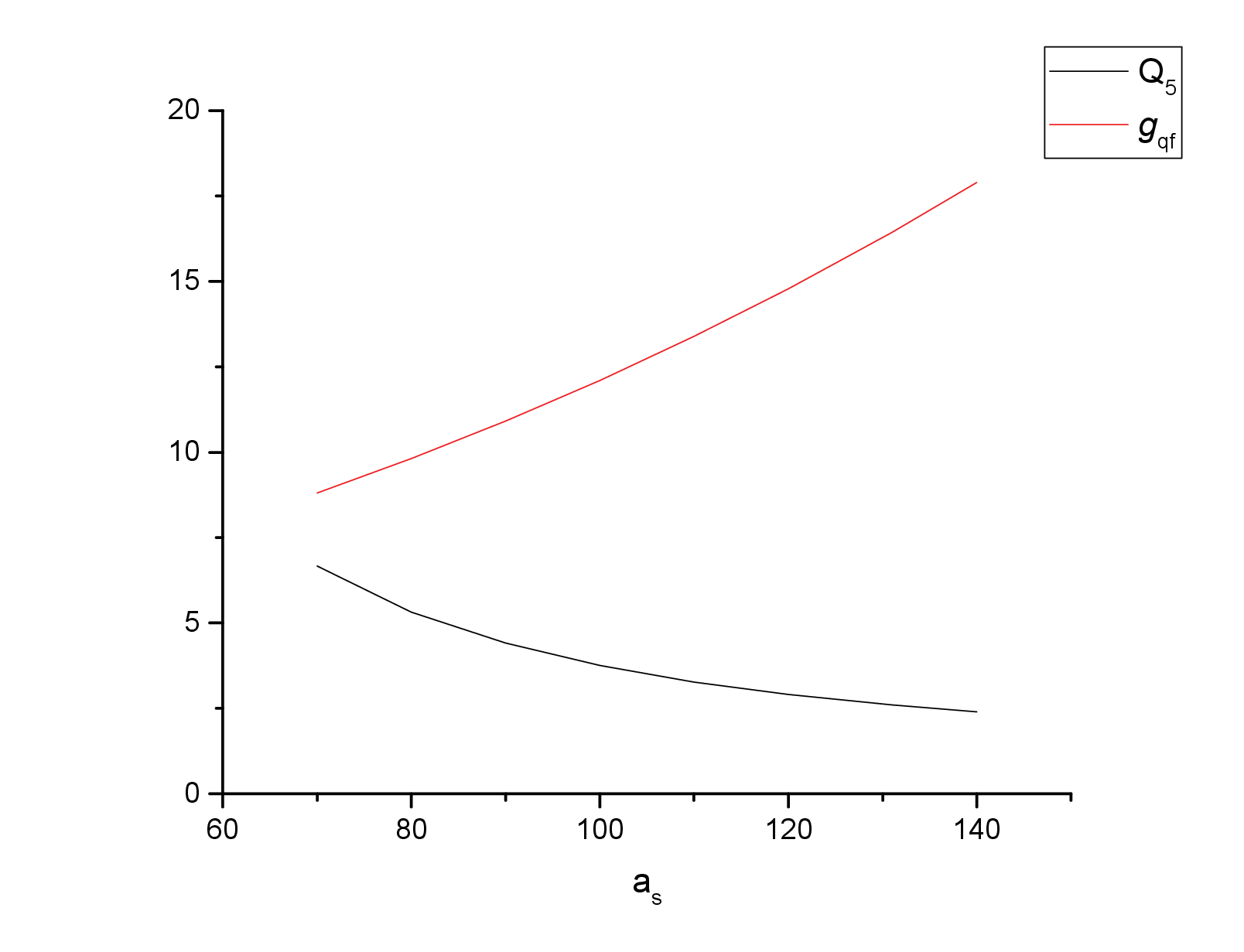}
\caption{(color online) The dependence of the dimensionless parameter $g_{%
\text{qf}}$ for the quantum fluctuations and the factor $Q_{5}$ on the
s-wave scattering length $a_{s}$, which is in unit of $a_{B}$, where $a_{B}$
is the Bohr radius. The atom number is $N=1000$.}
\label{fig1}
\end{figure}

The quantum fluctuations are important for the formation of stable droplets,
which is measured by the parameters $g_{\text{qf}}$ and $Q_{5}$. In Fig. (%
\ref{fig1}), we show the relation between them as the function of the $s$%
-wave scattering length $a_{s}$. It is shown that they have opposite
dependence on $a_{s}$.

In this paper, we consider the following composite trapping potential

\begin{equation}
V\left( \mathbf{r}\right) =\frac{1}{2}M\omega ^{2}\left( x^{2}+y^{2}\right)
+V_{0}\sin ^{2}\left( k_{L}z\right) ,
\end{equation}%
where $\omega $ is the trapping frequency of the symmetric harmonic
potential in the radial direction, which is taken to be $2\pi \times 44Hz$
in this paper. $V_{0}$\ is the depth of the optical lattice and $k_{L}$ is
the wavevector of the laser light.

The dynamical equation of the order parameter can be obtained from the
action principle \cite{pet02},

\begin{equation}
\delta \int_{t_{1}}^{t_{2}}Ldt=0,
\end{equation}%
where the Lagrangian $L$ is given by

\begin{equation}
L=N\int d\mathbf{r}\left\{ \frac{i\hbar }{2}\left[ \psi ^{\ast }\frac{%
\partial \psi }{\partial t}-\psi \frac{\partial \psi ^{\ast }}{\partial t}%
\right] \right\} -E.
\end{equation}%
The variation of $\psi ^{\ast }$ leads to the extended Gross-Pitaevskii
equation

\begin{equation}
i\hbar \frac{\partial \psi }{\partial t}=\left[ -\frac{\hbar ^{2}}{2m}%
\mathbf{\triangledown }^{2}+V+g\left\vert \psi \right\vert ^{2}+\Phi
_{dd}+g_{\text{qf}}\left\vert \psi \right\vert ^{3}\right] \psi .  \label{de}
\end{equation}

Convenient dimensionless parameters can be defined in terms of the harmonic
frequency $\omega $ and the corresponding oscillator length $l=\sqrt{\hbar
/M\omega }$. Using the dimensionless variables $\mathbf{\tilde{r}=r/}l$, $%
\tilde{a}_{s,dd}=a_{s,dd}/l$, $\tilde{t}=t\omega $, $\tilde{\psi}%
=l^{3/2}\psi $, $\tilde{V}_{0}=V_{0}/\hbar \omega $, $\tilde{k}_{L}=k_{L}l$,
the dynamical equation (\ref{de}) can be rewritten as

\begin{equation}
i\frac{\partial \tilde{\psi}}{\partial \tilde{t}}=\left[ -\frac{1}{2}\mathbf{%
\triangledown }^{2}+\tilde{V}+4\pi N\tilde{a}_{s}\left\vert \tilde{\psi}%
\right\vert ^{2}+\tilde{\Phi}_{dd}+\tilde{g}_{\text{qf}}\left\vert \tilde{%
\psi}\right\vert ^{3}\right] \tilde{\psi}  \label{de2}
\end{equation}%
where $\tilde{g}_{\text{qf}}=\frac{128}{3}\sqrt{\pi }N^{3/2}\sqrt{\tilde{a}%
_{s}^{5}}Q_{5}\left( \varepsilon _{dd}\right) $, and

\begin{equation}
\tilde{V}\left( \mathbf{\tilde{r}}\right) =\frac{1}{2}\left( \tilde{x}^{2}+%
\tilde{y}^{2}\right) +\tilde{V}_{0}\sin ^{2}\left( \tilde{k}_{L}\tilde{z}%
\right) ,
\end{equation}%
and

\begin{equation}
\tilde{\Phi}_{dd}=3N\tilde{a}_{dd}\int d\mathbf{\tilde{r}}^{^{\prime
}}\left\vert \psi \left( \mathbf{\tilde{r}}^{^{\prime }},\tilde{t}\right)
\right\vert ^{2}\frac{\left( 1-3\text{cos}^{2}\theta \right) }{\left\vert
\mathbf{\tilde{r}-\tilde{r}}^{^{\prime }}\right\vert ^{3}}.
\end{equation}

From now on we drop the over head $``\sim "$ with the dimensionless
variables understood. The ground states can be obtained with the method of
imaginary-time propagation of the dynamical equation (\ref{de2}). The term
of dipolar interaction can be treated with the help of convolution theorem
and Fourier transformation \cite{adh15}.

\section{Self-bound droplets}

To highlight the effects of an optical lattice on the properties of quantum
droplets, in this section we first consider the self-bound droplets without
an optical lattice. We consider the self-bound droplets along the
polarization direction,

\begin{equation}
V\left( \mathbf{r}\right) =\frac{1}{2}\left( x^{2}+y^{2}\right) ,
\end{equation}%
namely, without any trap along $z$-direction. So the stable ground states
obtained are self-bound by the two-body interactions and quantum
fluctuations.

\begin{figure}[tbp]
\centering\includegraphics  [width=8.6cm] {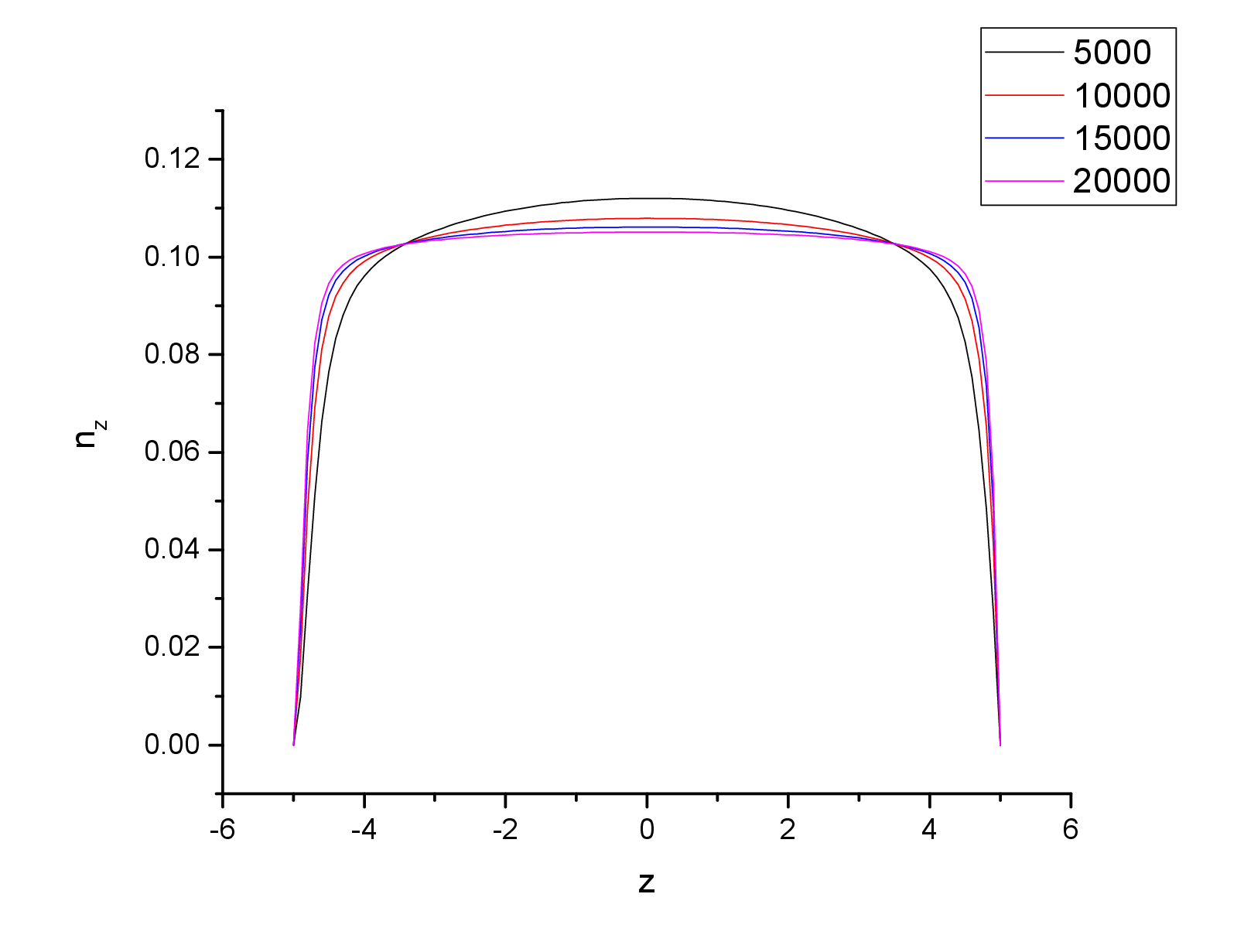}
\caption{(color online) The integrated density distribution $n_{z}$ of the
ground state for some atom numbers along $z$-direction. The $s$-wave
scattering length is taken to be $120a_{B}$, where $a_{B}$ is the
dimensionless Bohr radius in the unit of $l$.}
\label{fig2}
\end{figure}

\begin{figure}[tbp]
\centering\includegraphics  [width=8.6cm] {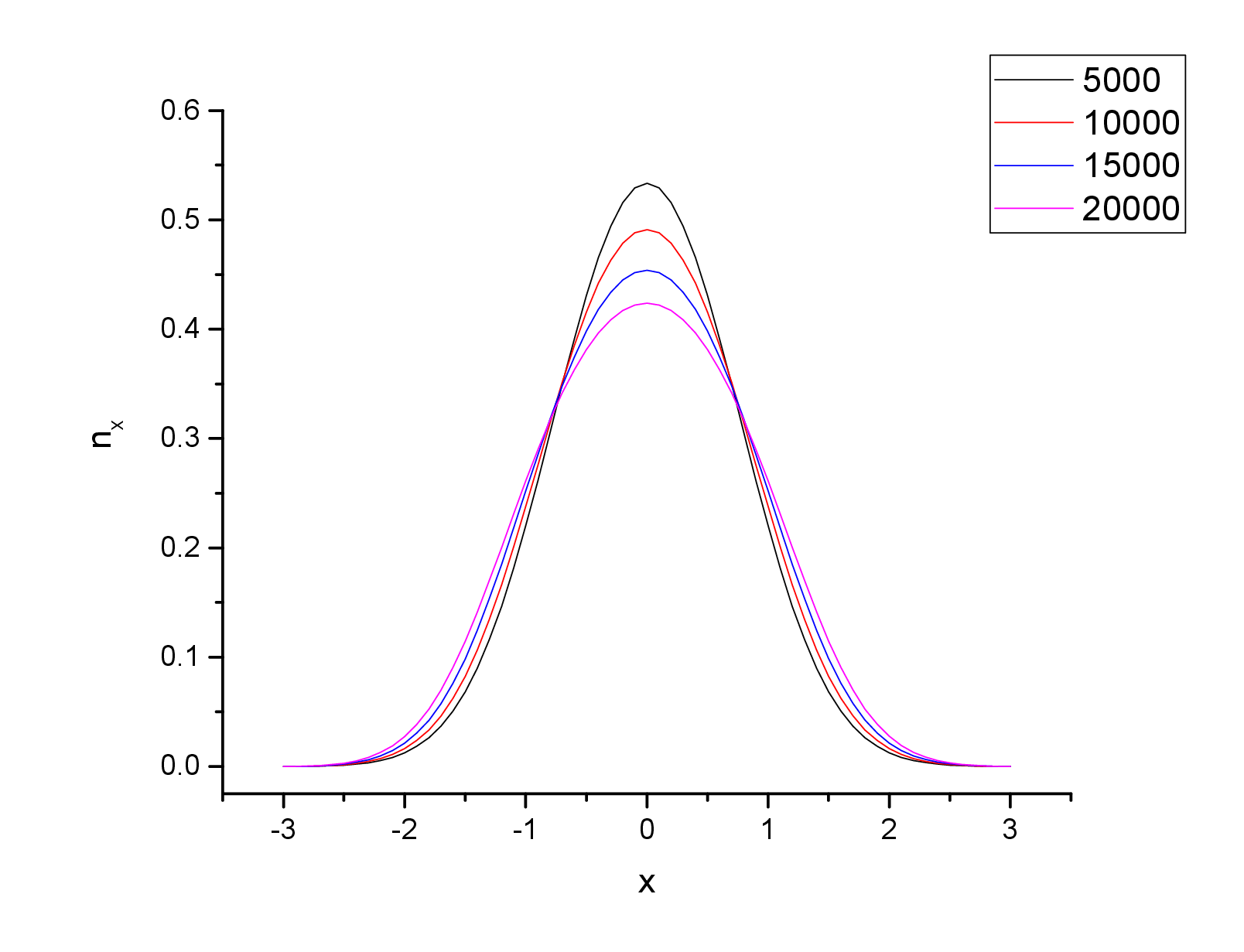}
\caption{(color online) The integrated density distribution $n_{x}$ of the
ground state for some atom numbers along $x$-direction. The $s$-wave
scattering length is the same with that in Fig. \protect\ref{fig2}.}
\label{fig3}
\end{figure}

In Fig. \ref{fig2} and \ref{fig3}, we show the density distributions for
various atom numbers along the $z$- and $x$-direction, respectively. The one
dimensional density distributions have been obtained by the integration of
the other two dimensions, for example, $n_{z}=\int \int \left\vert \psi
\right\vert ^{2}dxdy$. The similar definition is for the other density
distributions, such as $n_{x}$.

For $^{164}Dy$ atoms, the ground-state phase diagrams with or without the
harmonic trap have been investigated extensively. The parameters in Fig. \ref%
{fig2} and \ref{fig3} are chosen in the region of droplet. The density
platforms of the droplets along $z$-direction increase for the more atoms.
Clearly, the self-bound droplets have been obtained because the trap along $%
z $-direction is absent now. The density distributions along the radial
direction are still determined by the harmonic trap and the
contact-interaction.

\begin{figure}[tbp]
\centering\includegraphics  [width=8.6cm] {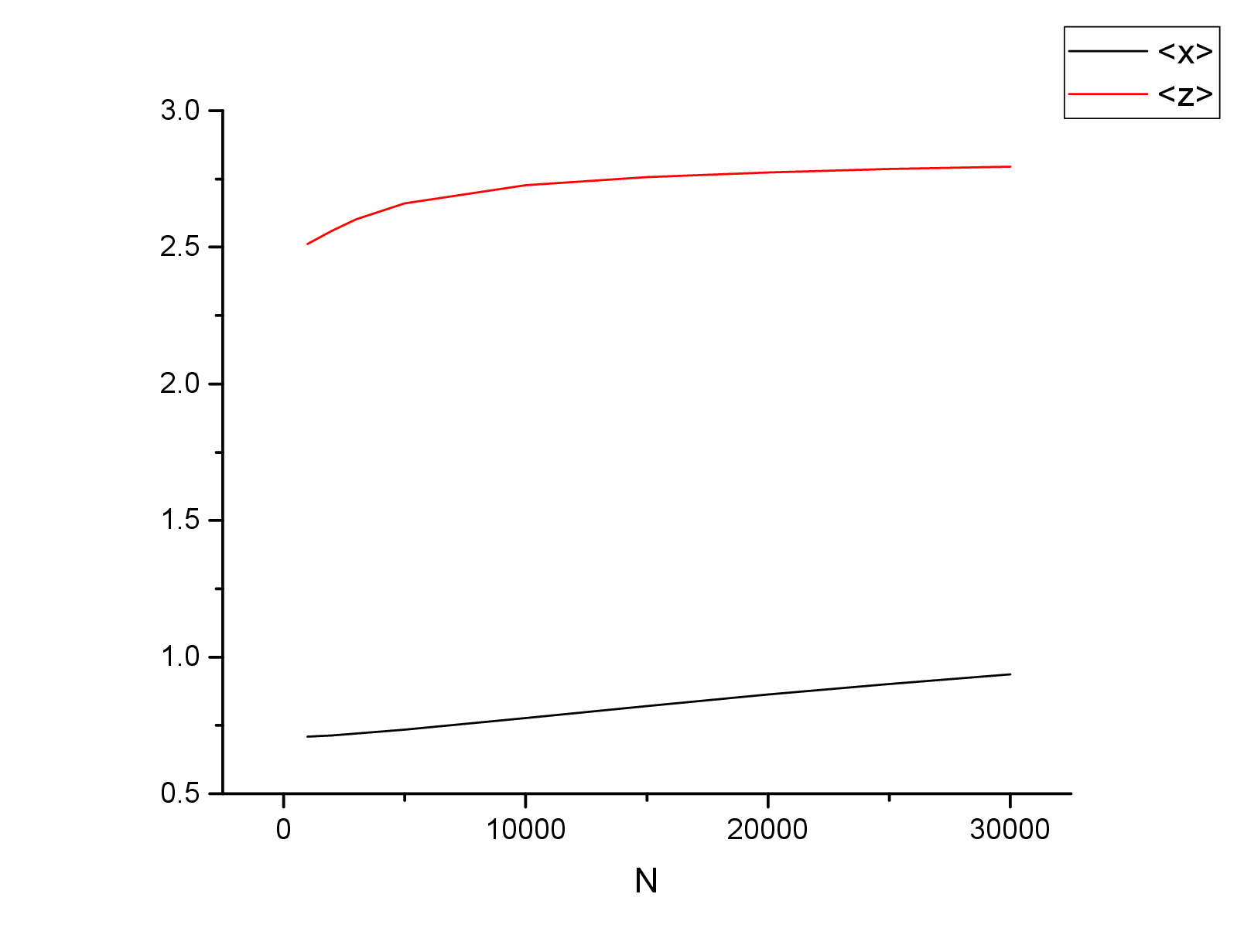}
\caption{(color online) The dependence of the exact widths along $x$- and $z$%
-directions of the ground state on the atom numbers. The $s$-wave scattering
length is the same with that in Fig. \protect\ref{fig2}.}
\label{fig4}
\end{figure}

\begin{figure}[tbp]
\centering\includegraphics  [width=8.6cm] {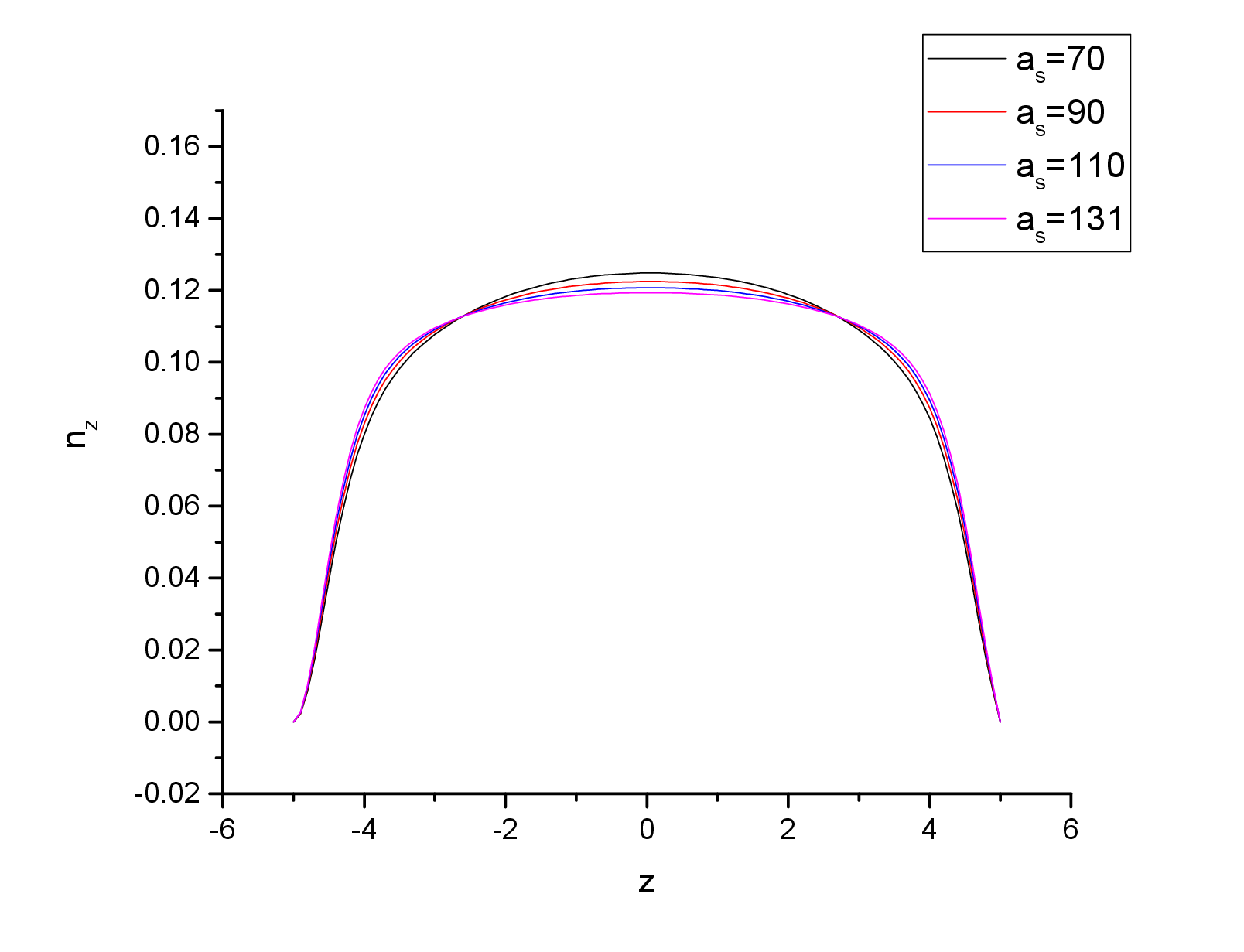}
\caption{(color online) The integrated density distribution $n_{z}$ of the
ground state for some $s$-wave scattering length $a_{s}$ along $z$%
-direction. The atom number is taken to be $1000$.}
\label{fig5}
\end{figure}

\begin{figure}[tbp]
\centering\includegraphics  [width=8.6cm] {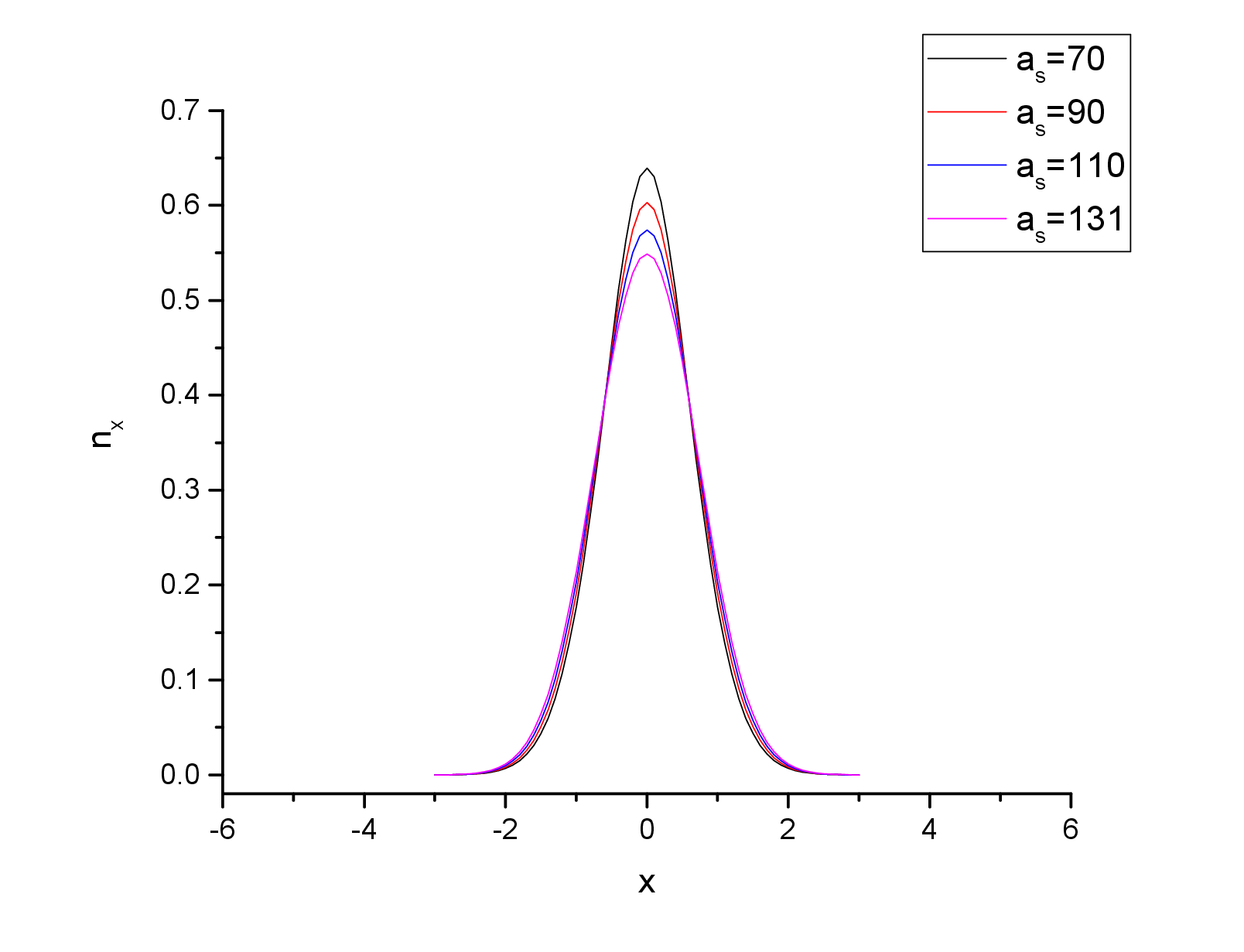}
\caption{(color online) The integrated density distribution $n_{x}$ of the
ground state for some $s$-wave scattering length $a_{s}$ along $x$%
-direction. The atom number is the same with that in Fig. \protect\ref{fig5}%
. }
\label{fig6}
\end{figure}

\begin{figure}[tbp]
\centering\includegraphics  [width=8.6cm] {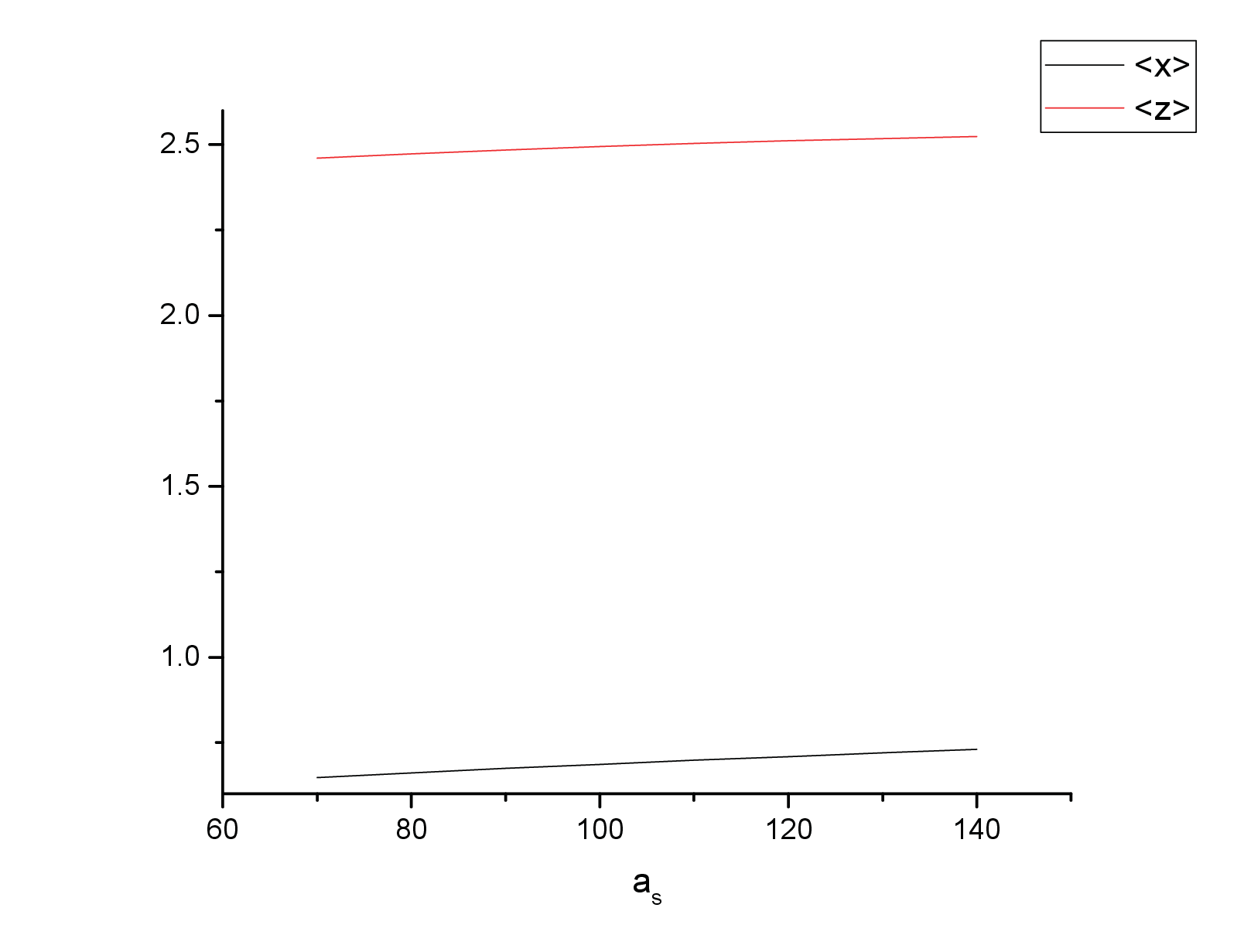}
\caption{(color online) The dependence of the exact widths along $x$- and $z$%
-directions of the ground state on the $s$-wave scattering length. The atom
number is the same with that in Fig. \protect\ref{fig5}.}
\label{fig7}
\end{figure}

The exact width scale of the atom cloud can be obtained in terms of
ground-state wavefunction with the definition $\left\langle z\right\rangle =%
\sqrt{\int z^{2}\left\vert \psi \right\vert ^{2}d\mathbf{r}}$. A similar
definition is for $\left\langle x\right\rangle $. In Fig. \ref{fig4}, we
show the exact widths along $x$- and $z$-directions for various atom number $%
N$. The density distributions along $z$-direction will be further flatten
for the more atoms and the exact width remains almost as it is. The exact
width along $x$-direction grows faster than that of $z$-direction.

In Fig. \ref{fig5} and \ref{fig6}, we show the density distributions for
various $s$-wave scattering length $a_{s}$ along the $z$- and $x$-direction,
respectively. For the given total atom number, the peak density in Fig. \ref%
{fig5} decreases with the growth of $a_{s}$. For the stronger repulsive
contact-interaction, the relatively weaker attraction between the dipoles
aligned along $z$-direction results in the decrease of the peak density. In
Fig. \ref{fig6}, the peak density also decreases with the growth of $a_{s}$.
The stronger repulsive contact-interaction expands the atom cloud agaist the
relatively weaker confining trap along the radial direction and the peak
density reduces.

\begin{figure}[tbp]
\centering\includegraphics  [width=8.6cm] {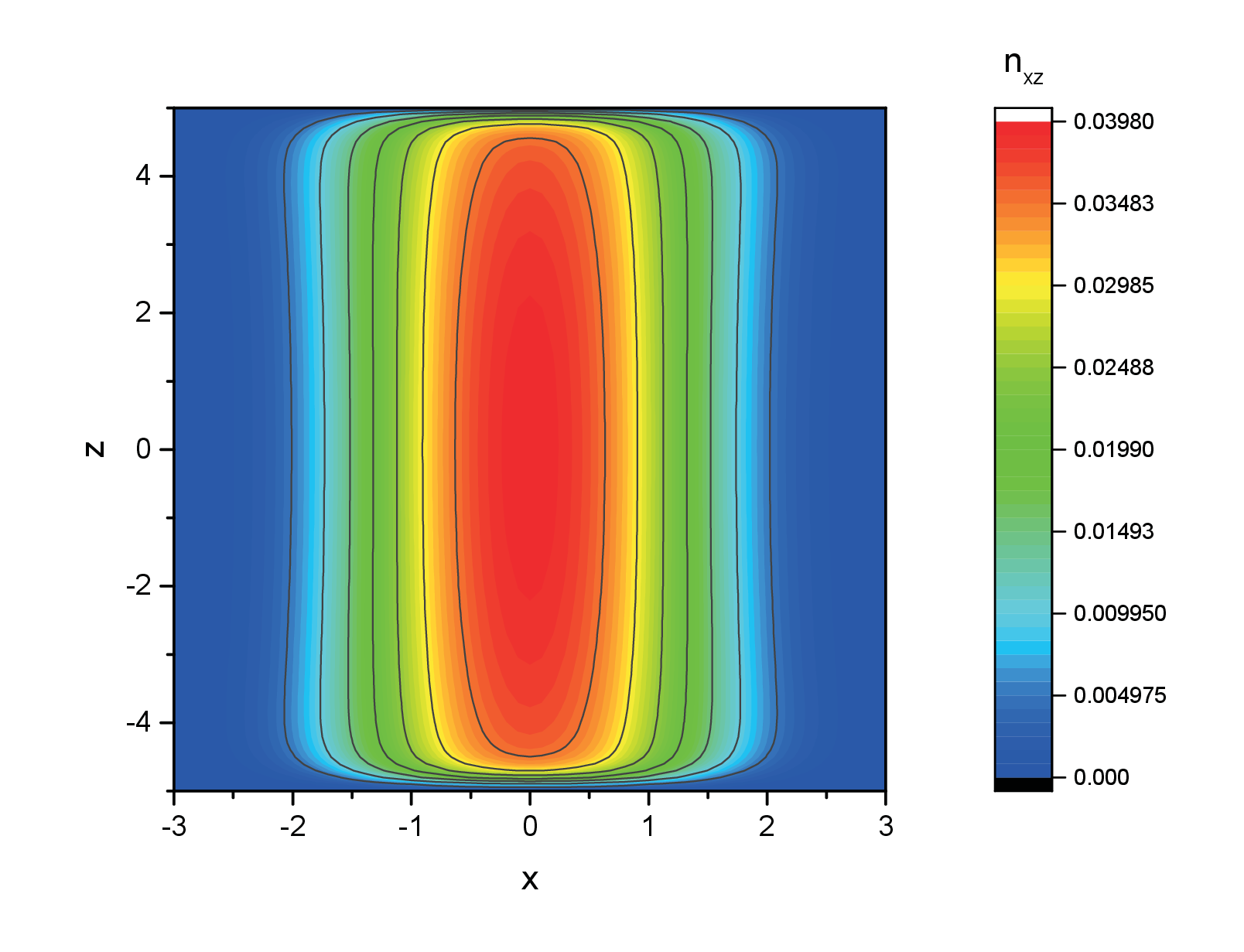}
\caption{(color online) The integrated density distribution $n_{xz}$ of the
ground state in $x$\text{-}$z$ plane. The $s$-wave scattering length is
taken to be $120a_{B}$. The atom number is taken to be $1000$.}
\label{fig8}
\end{figure}

\begin{figure}[tbp]
\centering\includegraphics  [width=8.6cm] {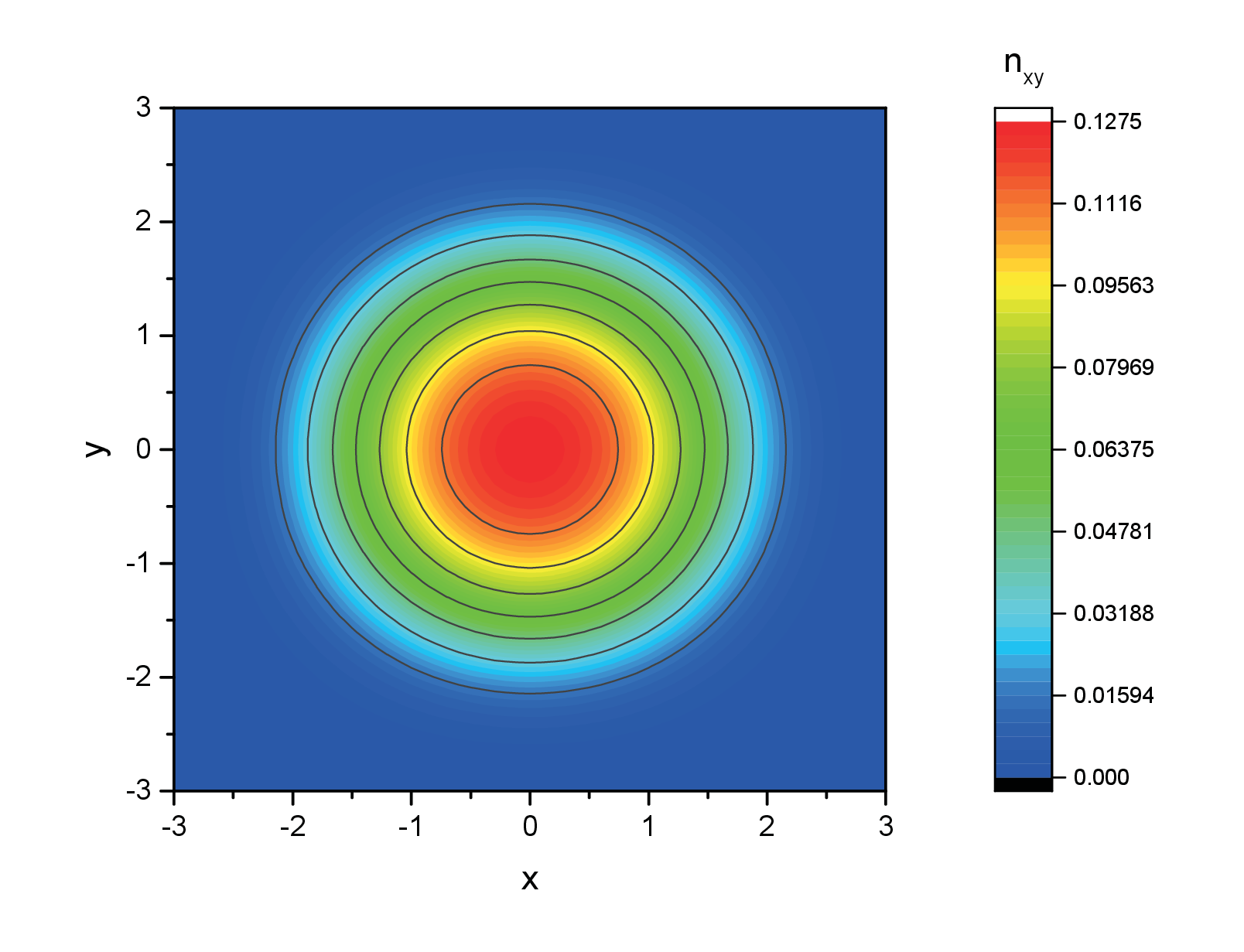}
\caption{(color online) The integrated density distribution $n_{xy}$ of the
ground state in $x$-$y$ plane. The parameters are taken to be the same with
those in Fig. \protect\ref{fig8}.}
\label{fig9}
\end{figure}

In Fig. \ref{fig7}, we show the exact widths along $z$- and $x$-directions
for various $s$-wave scattering length $a_{s}$. The width along $z$%
-direction grows slower than that along $x$-direction due to the attraction
between the dipoles.

In Fig. \ref{fig8} and \ref{fig9}, we show the density distributions in $x-z$
and $x-y$ plane, respectively. The two dimensional distributions have been
obtained by the integration of the other one dimension, for example, $%
n_{x,y}=\int \left\vert \psi \right\vert ^{2}dz$. The distribution in $x-y$
plane is mainly determined by the trap and the repulsive
contact-interaction. But the distribution in $x-z$ plane is determined by
the two-body interactions and quantum fluctuations.

\section{Self-bound droplets in an optical lattice}

Next, we consider the effects of an optical lattice on the properties of the
self-bound droplets, namely,

\begin{equation}
V\left( \mathbf{r}\right) =\frac{1}{2}\left( x^{2}+y^{2}\right) +V_{0}\sin
^{2}\left( k_{L}z\right) .
\end{equation}

\begin{figure}[tbp]
\centering\includegraphics  [width=8.6cm] {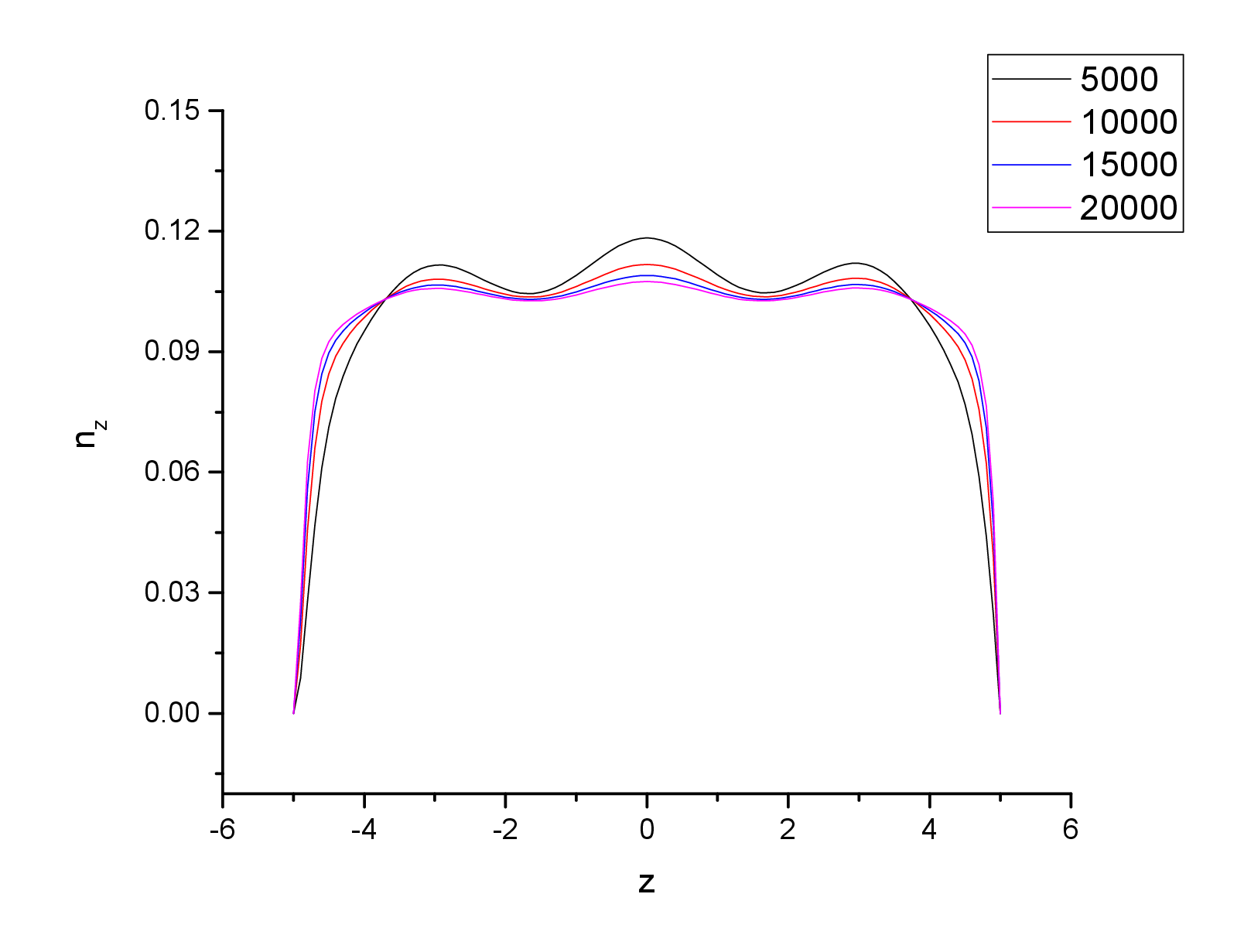}
\caption{(color online) The integrated density distribution $n_{z}$ of the
ground state for some atom numbers along $z$-direction. The $s$-wave
scattering length is the same with that in Fig. \protect\ref{fig2}. The
dimensionless depth $V_{0}$ and wavevector $k_{L}$ of the optical lattice
are both taken to be 1.}
\label{fig10}
\end{figure}

\begin{figure}[tbp]
\centering\includegraphics  [width=8.6cm] {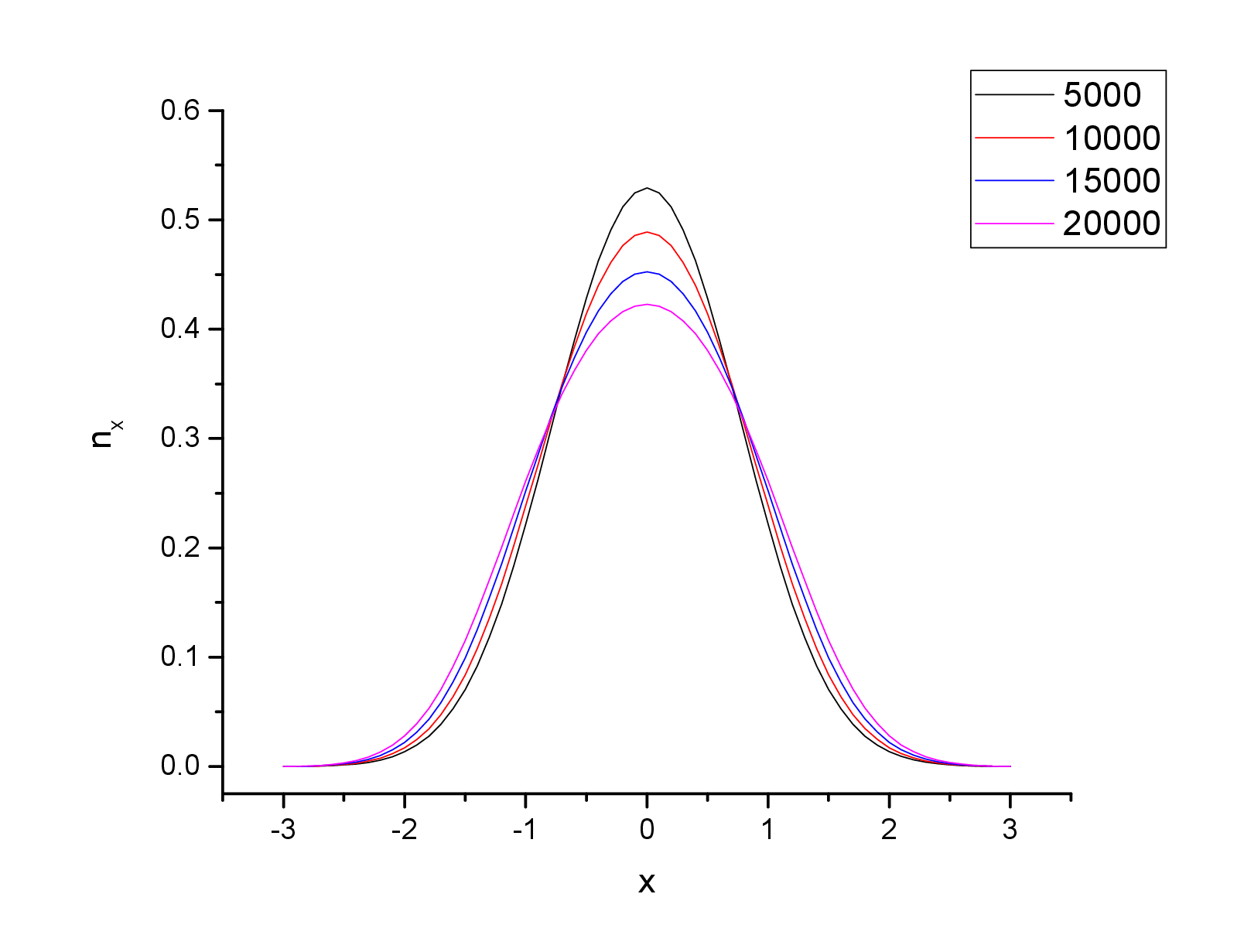}
\caption{(color online) The integrated density distribution $n_{x}$ of the
ground state for some atom numbers along $x$-direction. The parameters are
taken to be the same with those in Fig. \protect\ref{fig10}.}
\label{fig11}
\end{figure}

We want to emphasize here that the depth of the optical lattice is weak
considered in this paper. For the strong optical lattice, the atoms will be
localized at the bottom of the lattice. The system has to be described by
the well-known Bose-Hubbard model with the wavefunction expanding by the
local Wannier funcions.

In Fig. \ref{fig10} and \ref{fig11}, we show the density distributions for
various atom number along $z$- and $x$-directions, respectively. The density
distributions along the radial direction are still determined by the trap
and the contact interaction. It is shown in Fig. \ref{fig10} that with the
increase of atoms, the density distributions along $z$-direction tend to be
those without the optical lattice in Fig. \ref{fig2}. The optical lattice
has only the effects of modulating the droplets. For the more atoms, the
interactions and quantum fluctuations, which are proportional to atom
number, are enhanced and the effects of the optical lattice become weaker
and weaker. So we still call them the self-bound droplets, because the
confining harmonic trap along $z$ is absent now. The stable ground states
are still determined by the two-body interactions and quantum fluctuations.

\begin{figure}[tbp]
\centering\includegraphics  [width=8.6cm] {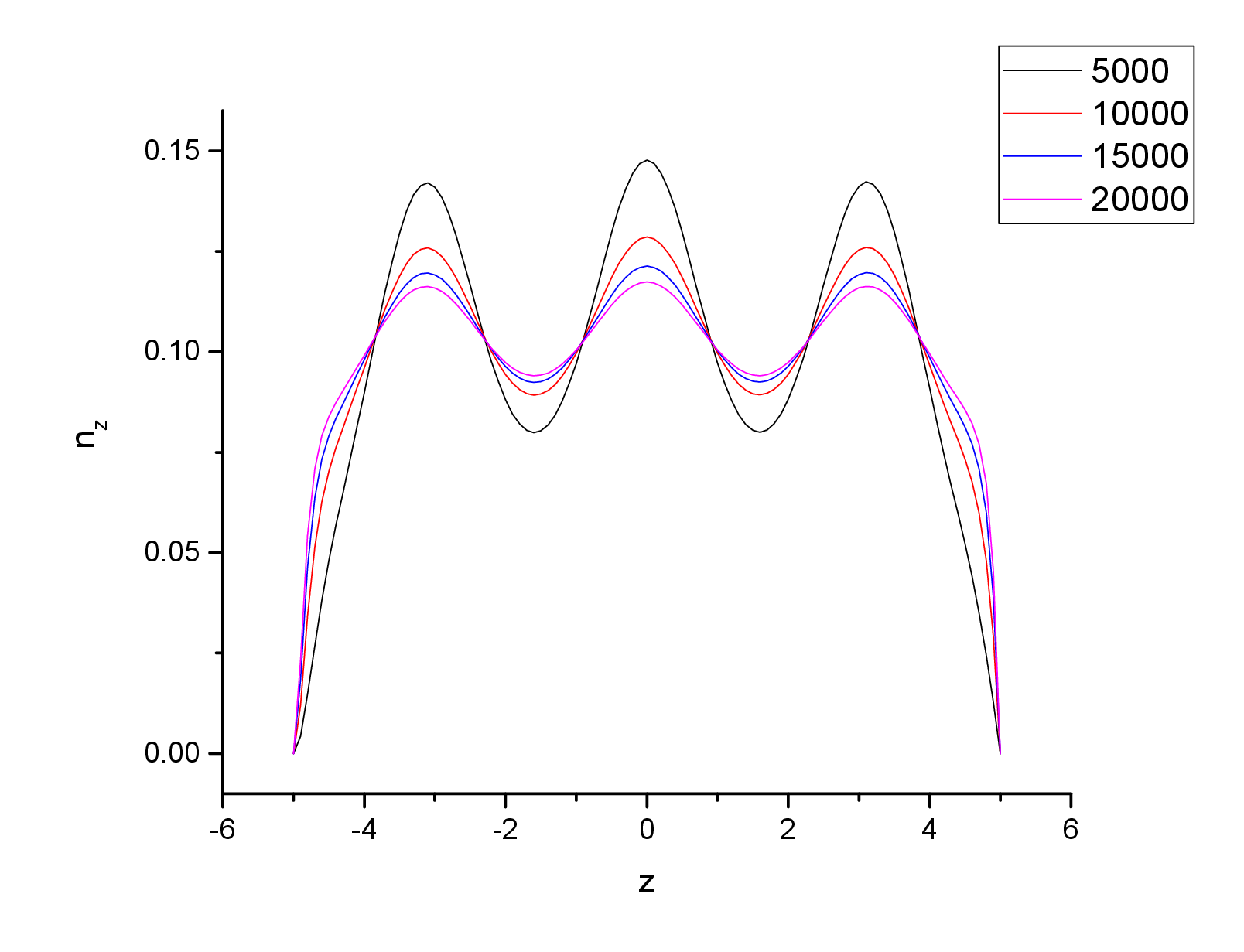}
\caption{(color online) The integrated density distribution $n_{z}$ of the
ground state for some atom numbers along $z$-direction. The parameters are
taken to be the same with those in Fig. \protect\ref{fig10}, except the
dimensionless depth $V_{0}$ of the optical lattice is taken to be 5.}
\label{fig12}
\end{figure}

To confirm this, we consider the deeper optical lattice with the depth $%
V_{0} $ five times of that in Fig. \ref{fig10}. It is shown in Fig. \ref%
{fig12} that the optical lattice still modulates only the droplets and the
system behaves as that without the optical lattice for the larger atom
number.

\begin{figure}[tbp]
\centering\includegraphics  [width=8.6cm] {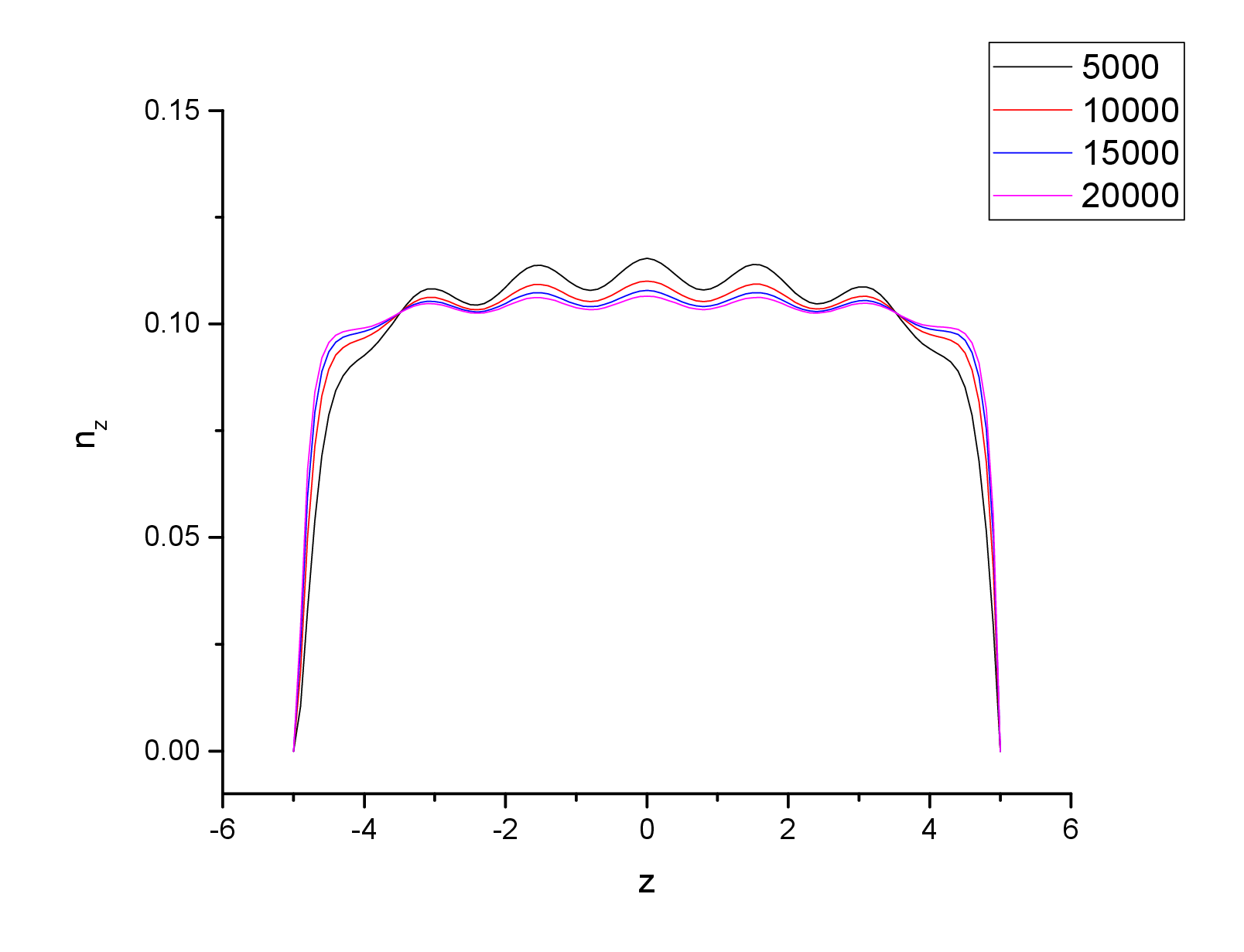}
\caption{(color online) The integrated density distribution $n_{z}$ of the
ground state for some atom numbers along $z$-direction. The parameters are
taken to be the same with those in Fig. \protect\ref{fig10}, except the
dimensionless wavevector $k_{L}$ is taken to be 2.}
\label{fig13}
\end{figure}

\begin{figure}[tbp]
\centering\includegraphics  [width=8.6cm] {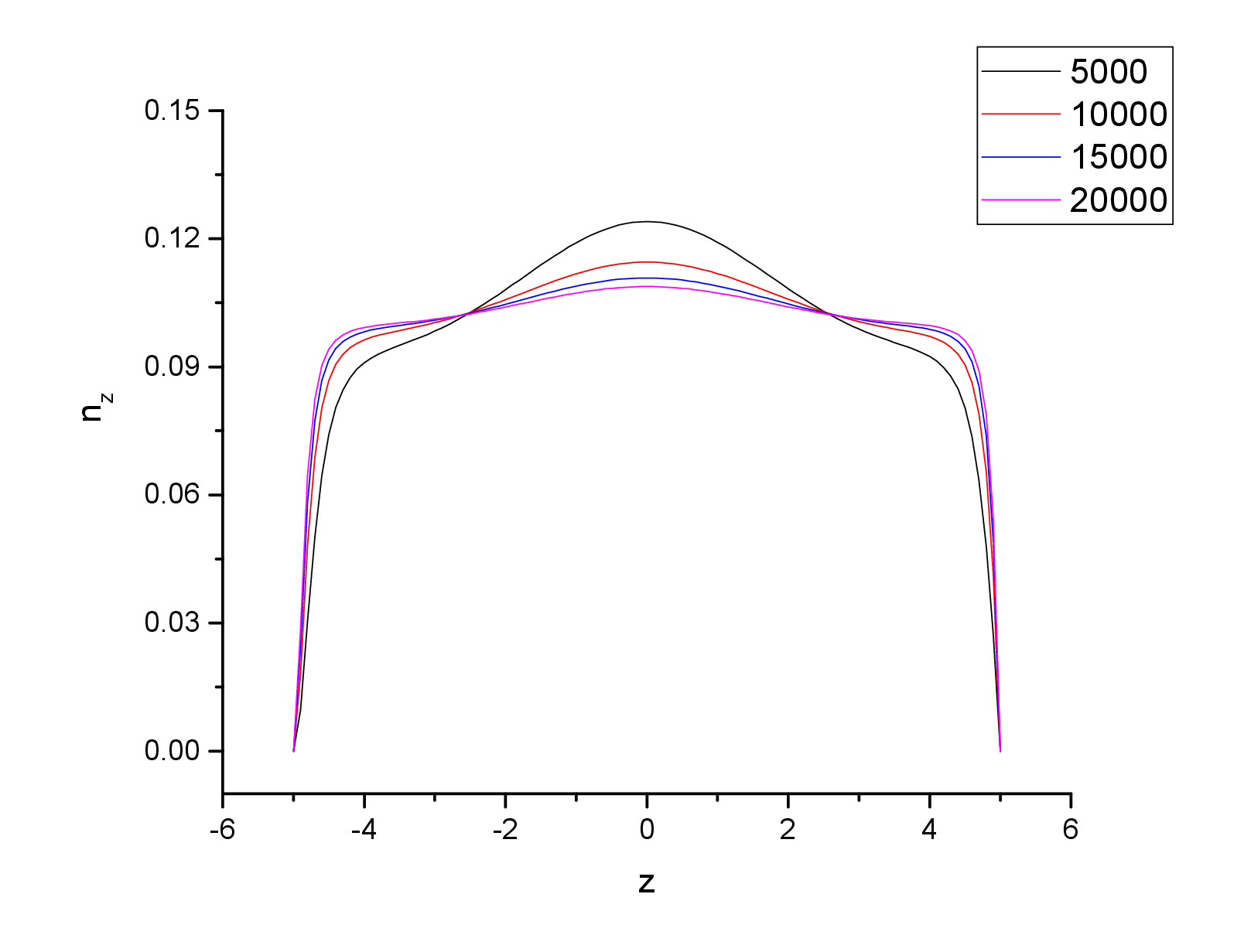}
\caption{(color online) The integrated density distribution $n_{z}$ of the
ground state for some atom numbers along $z$-direction. The parameters are
taken to be the same with those in Fig. \protect\ref{fig10}, except the
dimensionless wavevector $k_{L}$ is taken to be 0.5.}
\label{fig14}
\end{figure}

We also consider the effects of the wavevector $k_{L}$ on the droplets. We
show the density distributions along $z$-direction in Fig. \ref{fig13} and %
\ref{fig14} for the wavevector being twice and half of that in Fig. \ref%
{fig10}, respectively. It is shown again that the optical lattice only
modulates the droplets and the system tends to that without the optical
lattice for the larger atom number.

\begin{figure}[tbp]
\centering\includegraphics  [width=8.6cm] {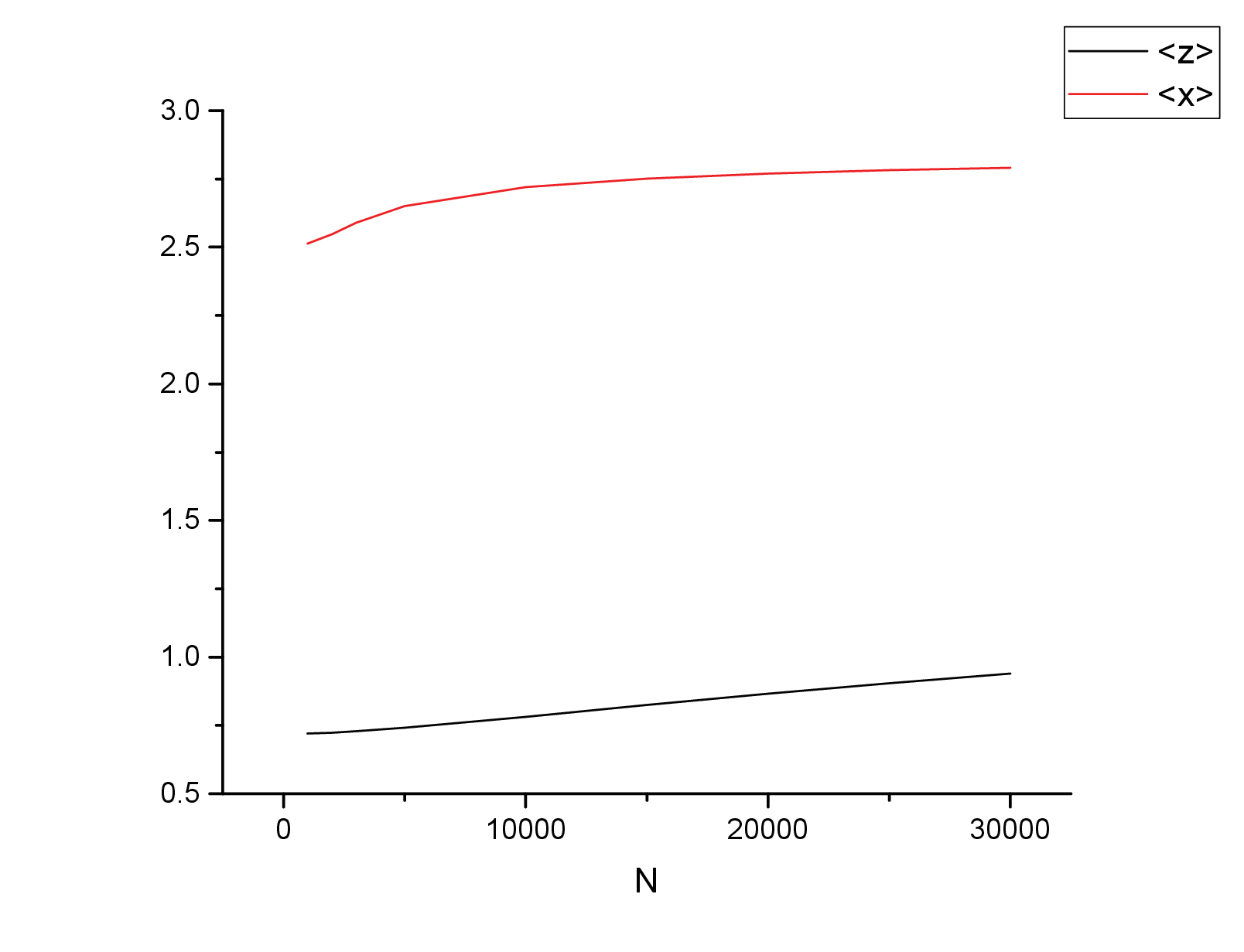}
\caption{(color online) The dependence of the exact widths along $x$- and $z$%
-directions of the ground state on the atom numbers. The parameters are
taken to be the same with those in Fig. \protect\ref{fig10}.}
\label{fig15}
\end{figure}

In Fig. \ref{fig15}, we show the exact widths along $z$- and $x$-directions
for various atom number $N$ in the presence of the optical lattice. The
density distributions along $z$-direction have been further flatten for the
more atoms and the exact width remains almost as it is. The width along $z$%
-direction grows slower than that along $x$-direction.

\begin{figure}[tbp]
\centering\includegraphics  [width=8.6cm] {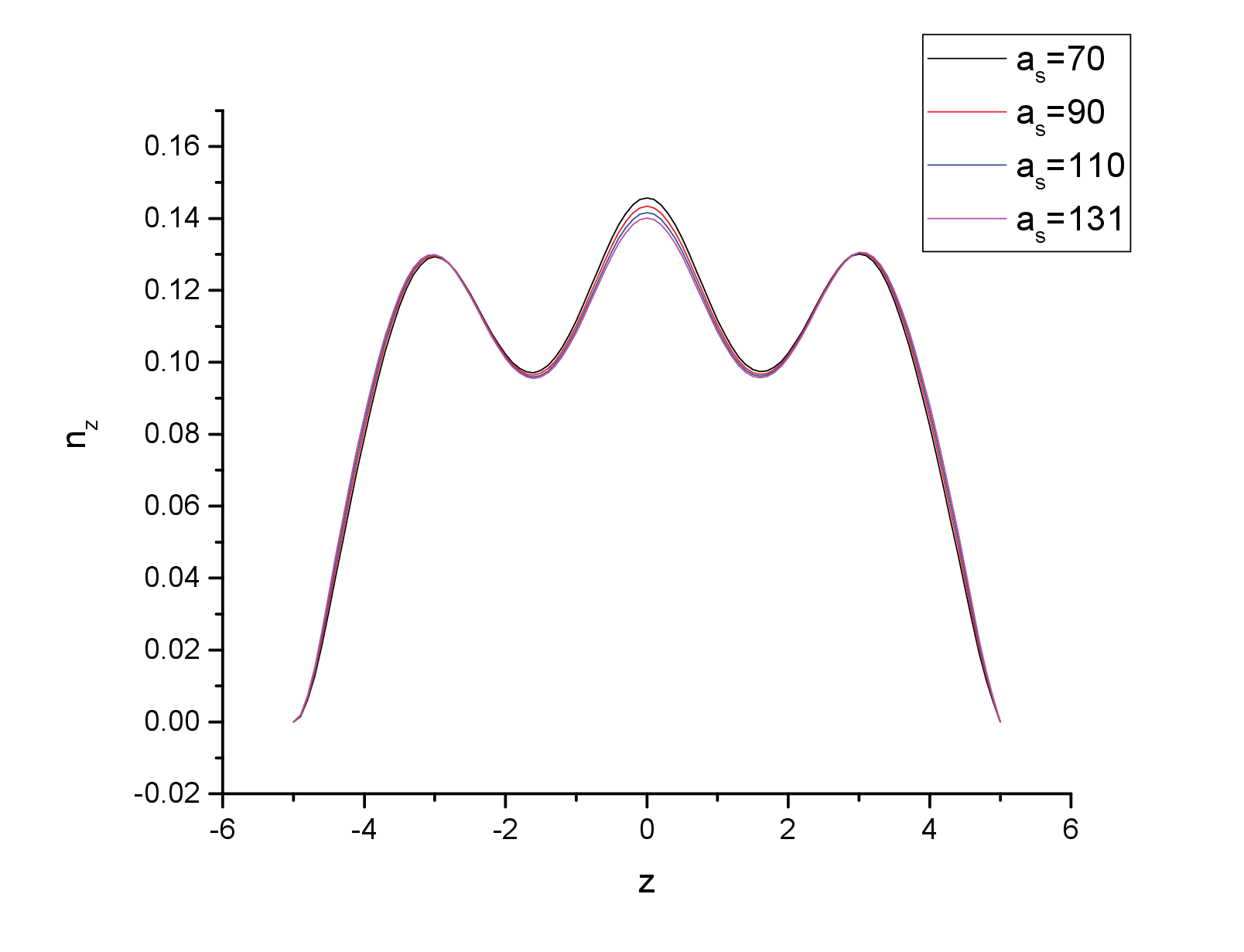}
\caption{(color online) The integrated density distribution $n_{z}$ of the
ground state for some $s$-wave scattering length $a_{s}$ along $z$%
-direction. The atom number is taken to be $1000$. The dimensionless depth $%
V_{0}$ and wavevector $k_{L}$ of the optical lattice are both taken to be 1.}
\label{fig16}
\end{figure}

\begin{figure}[tbp]
\centering\includegraphics  [width=8.6cm] {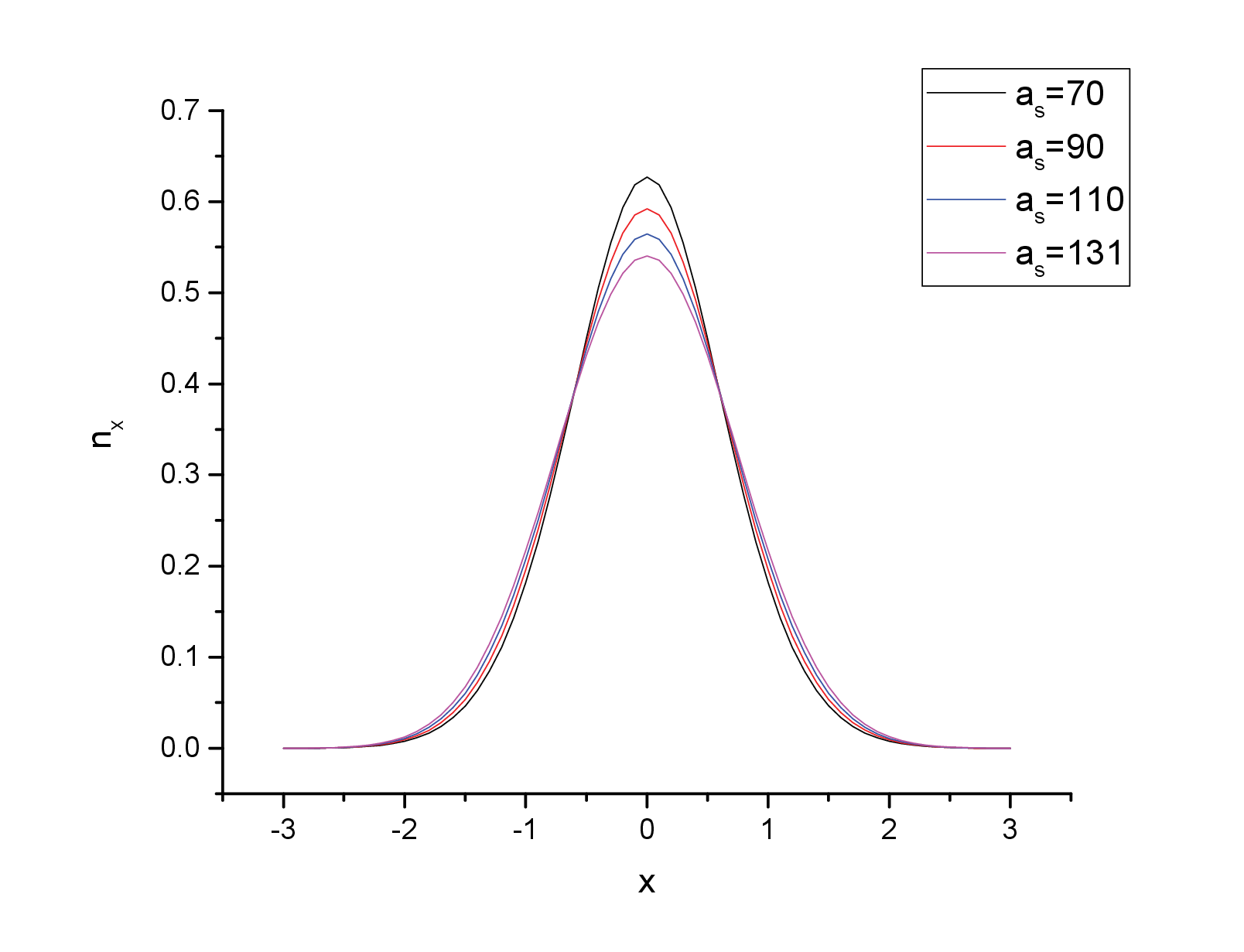}
\caption{(color online) The integrated density distribution $n_{x}$ of the
ground state for some $s$-wave scattering length $a_{s}$ along $x$%
-direction. The parameters are taken to be the same with those in Fig.
\protect\ref{fig16}.}
\label{fig17}
\end{figure}

In Fig. \ref{fig16} and \ref{fig17}, we show the density distributions for
various $s$-wave scattering length $a_{s}$ along the $z$- and $x$%
-directions, respectively. The stronger contact-interaction will expand the
atom cloud and reduce the peak density along radial direction, as that
without the optical lattice. But the stronger contact-interaction has lesser
effects on the density distribution along $z$-direction, except reducing the
peak density at the extreme points of the density distribution.

\begin{figure}[tbp]
\centering\includegraphics  [width=8.6cm] {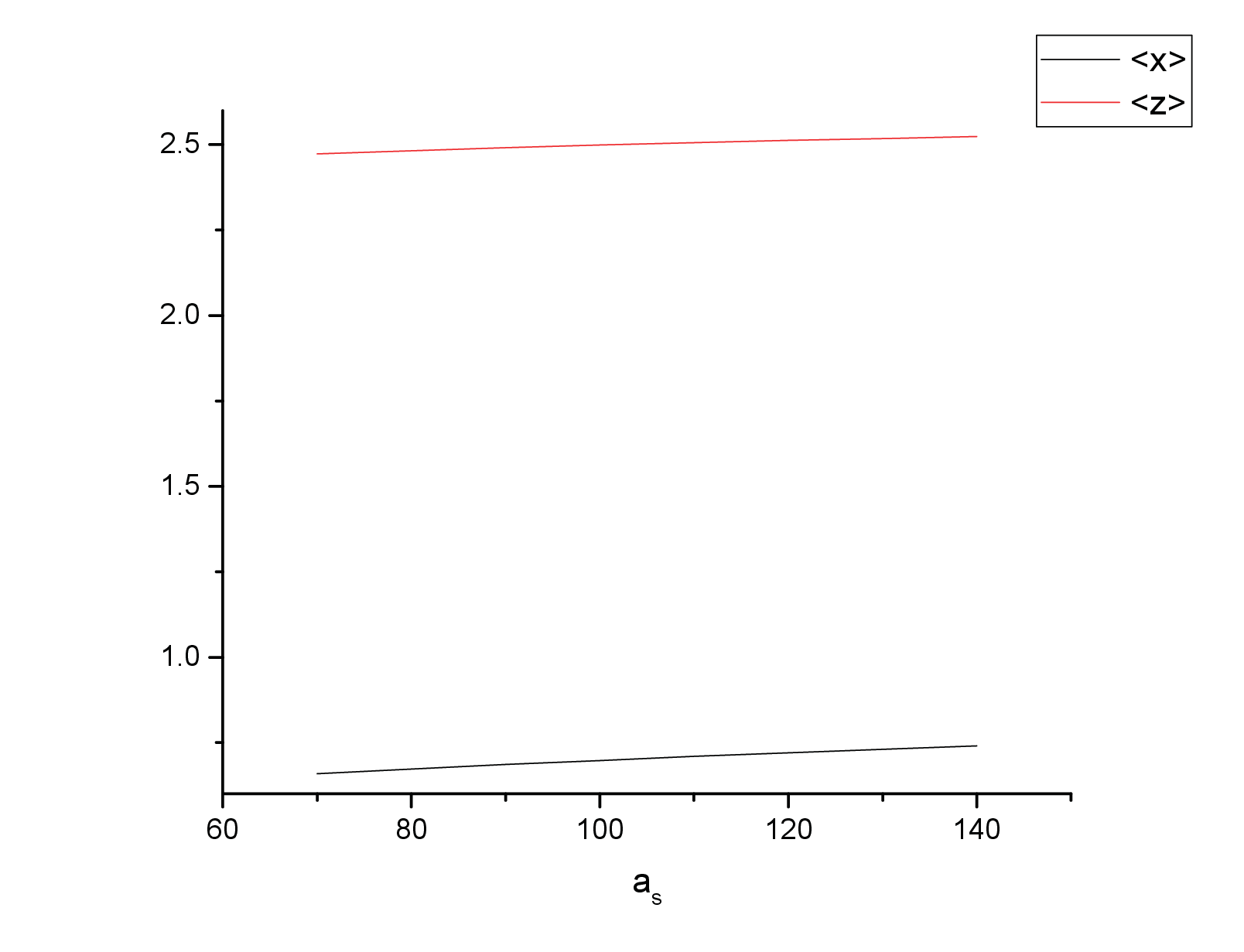}
\caption{(color online) The dependence of the exact widths along $x$- and $z$%
-directions of the ground state on the $s$-wave scattering length. The
parameters are taken to be the same with those in Fig. \protect\ref{fig16}.}
\label{fig18}
\end{figure}

In Fig. \ref{fig18}, we show the exact widths along $x$- and $z$-directions
for various $s$-wave scattering length $a_{s}$. The widths along the $x$-
and $z$-direction are almost the same with those without the optical
lattice. It confirms again that the optical lattice only modulates the
droplets. The formation of droplets is determined by the atomic interactions
and quantum fluctuations.

\begin{figure}[tbp]
\centering\includegraphics  [width=8.6cm] {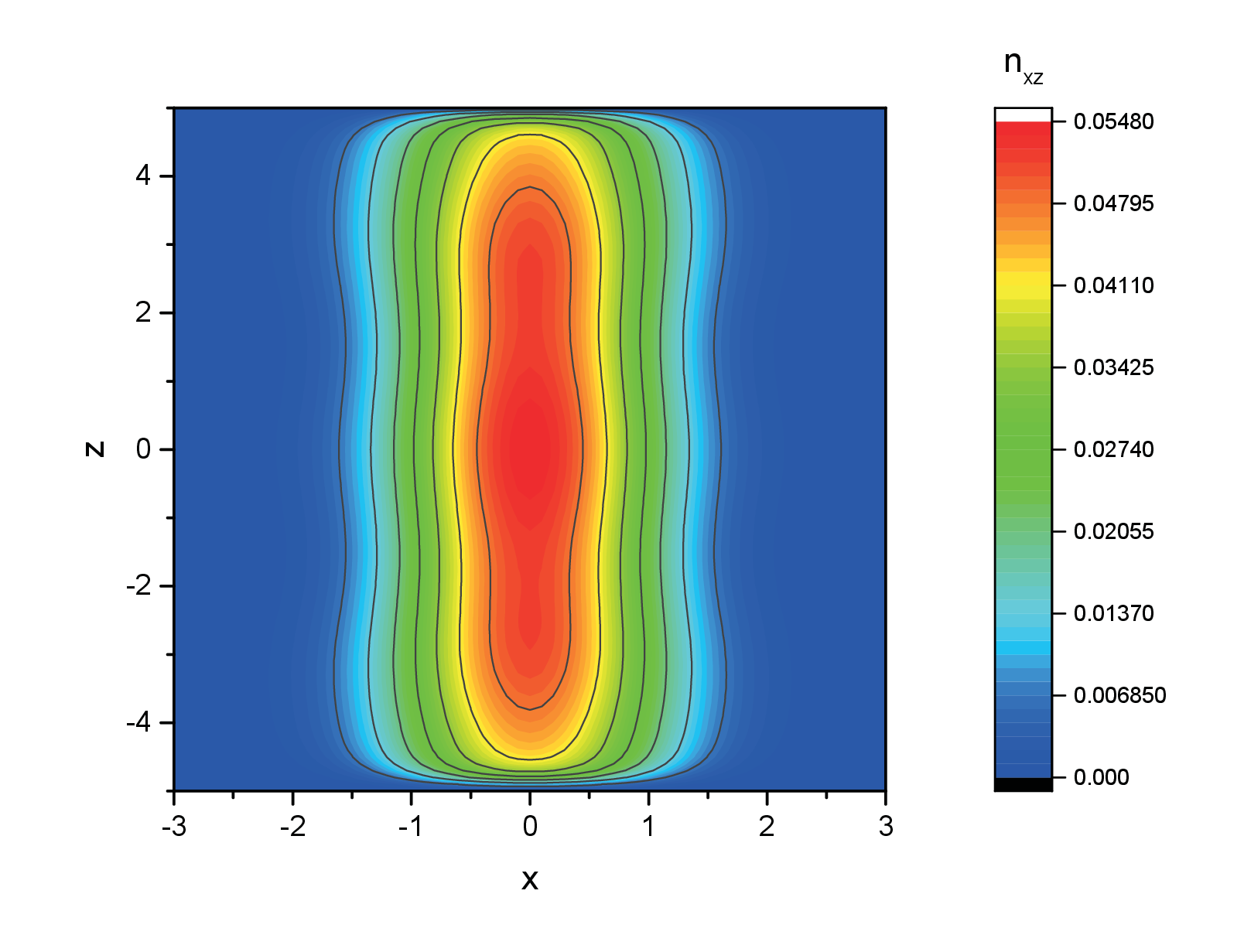}
\caption{(color online) The integrated density distribution $n_{xz}$ of the
ground state in the $x$-$z$ plane. The $s$-wave scattering length is taken
to be $120a_{B}$. The atom number is taken to be $1000$. The dimensionless
depth $V_{0}$ and wavevector $k_{L}$ of the optical lattice are both taken
to be 1.}
\label{fig19}
\end{figure}

\begin{figure}[tbp]
\centering\includegraphics  [width=8.6cm] {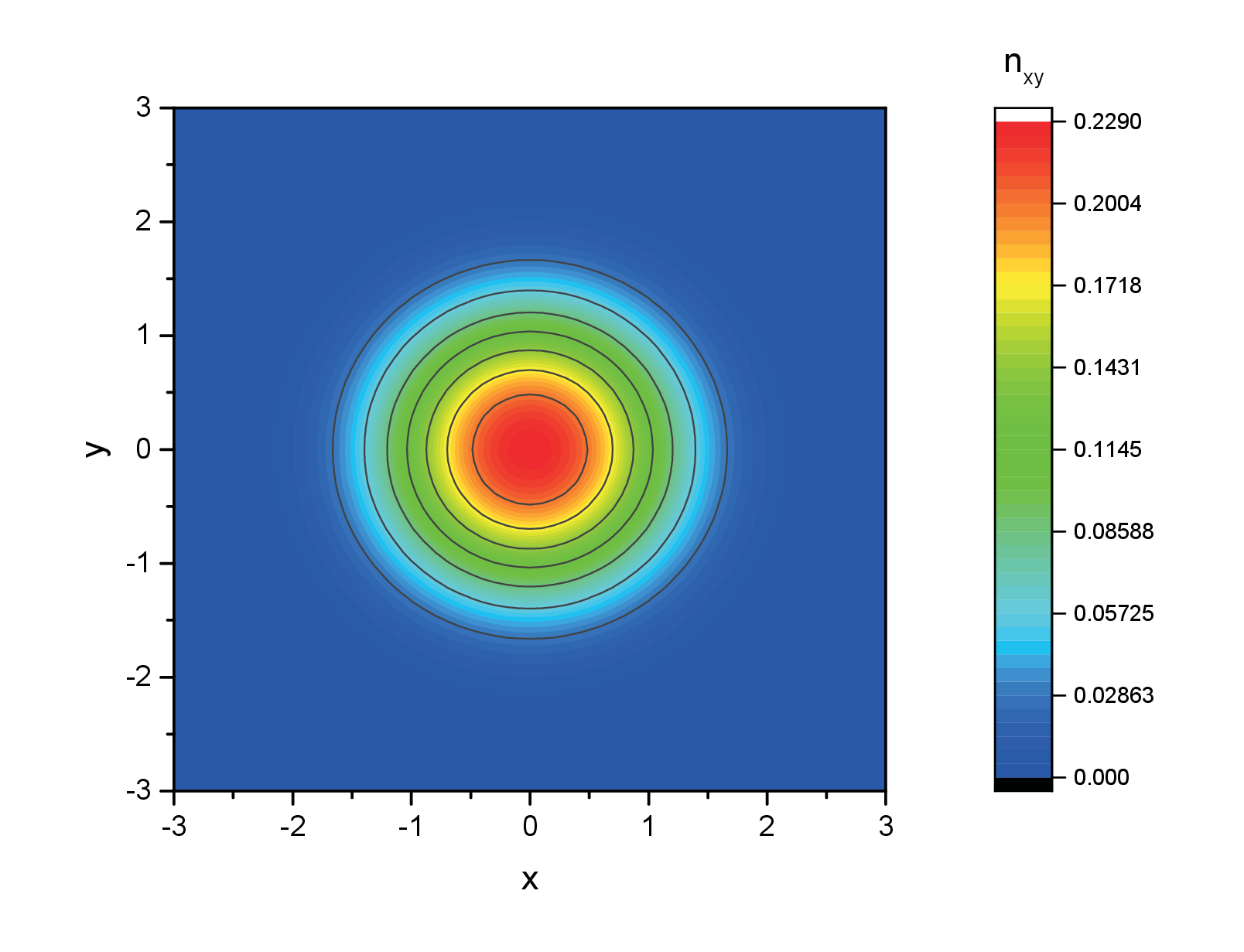}
\caption{(color online) The integrated density distribution $n_{xy}$ of the
ground state in the $x$-$y$ plane. The parameters are taken to be the same
with those in Fig. \protect\ref{fig19}.}
\label{fig20}
\end{figure}

In Fig. \ref{fig19} and \ref{fig20}, we show the density distributions in
the $x-z$ and $x-y$\ plane, respectively. A periodically-modulated quantum
droplet along the polarization direction is observed in Fig. \ref{fig19}.
The density distribution in $x-y$ plane is similar with that without the
optical lattice.

\section{Conclusion and some remark}

In summary, we have studied the properties of self-bound dipolar droplets
modulated by a shallow optical lattice. The properties of the quantum
droplets are investigated extensively. For the shallow enough optical
lattice, our results agree well with those without the optical lattice.

In the numerical evolution of the imaginary dynamical equation we have
adopted the zero boundary condition. We have also done the simulations with
the periodic boundary condition along $z$-direction, where the similar
results have been obtained.

Because we have adopted the zero boundary condition, our model is also
suitable for the finite-size system with two baffles blocking at both ends
along $z$-direction. The potential is still zero in the finite length scale
along $z$-direction except the optical lattice potential. If we broaden the
finite length scale along $z$-direction, the similar results can be obtained.

\section{Acknowledgment}

This work was supported by the National Natural Science Foundation of China,
under Grants No. 12047501.

\end{document}